\documentclass[aps,prx,preprint,showpacs,floatfix,superscriptaddress]{revtex4-1}

\usepackage{hyperref} 

\usepackage[utf8]{inputenc}
\usepackage{amsmath,amssymb,amstext}
\usepackage{braket}
\usepackage{amsthm}
\usepackage{dsfont}
\usepackage{color}

\def\sout{\bgroup\markoverwith
{\textcolor{red}{\rule[0.5ex]{2pt}{0.5pt}}}\ULon}
\def\be{\begin{equation}}
\def\ee{\end{equation}}
\def\bes{\begin{equation*}}
\def\ees{\end{equation*}}
\def\bea{\begin{eqnarray}}
\def\eea{\end{eqnarray}}
\def\beas{\begin{eqnarray*}}
\def\eeas{\end{eqnarray*}}
\def\bal#1\eal{\begin{align}#1\end{align}}
\def\bals#1\eals{\begin{align*}#1\end{align*}}

\usepackage[normalem]{ulem}

\usepackage{braket}

\begin{document}
\title{Emergence of Anyons on the Two-Sphere in Molecular Impurities}

\author{M. Brooks}
\email{morris.brooks@ist.ac.at}
\affiliation{IST Austria (Institute of Science and Technology Austria), Am Campus 1, 3400 Klosterneuburg, Austria}
\author{M. Lemeshko}
\affiliation{IST Austria (Institute of Science and Technology Austria), Am Campus 1, 3400 Klosterneuburg, Austria}
\author{D. Lundholm}
\affiliation{Department of Mathematics, Uppsala University - Box 480, SE-751 06 Uppsala, Sweden}
\author{E. Yakaboylu}
\email{enderalp.yakaboylu@ist.ac.at}
\affiliation{IST Austria (Institute of Science and Technology Austria), Am Campus 1, 3400 Klosterneuburg, Austria}
\affiliation{Max Planck Institute of Quantum Optics, 85748 Garching, Germany}

\date{\today}

\begin{abstract}

Recently it was shown that anyons on the two-sphere naturally arise from a system of molecular impurities exchanging angular momentum with a many-particle bath (Phys. Rev. Lett. 126, 015301 (2021)). 
Here we further advance this approach and rigorously demonstrate that in the experimentally realized regime the lowest spectrum of two linear molecules immersed in superfluid helium corresponds to the spectrum of two anyons on the sphere. 
We develop the formalism within the framework of the recently experimentally observed angulon quasiparticle.

\end{abstract}

\maketitle

\section{Introduction}

The discovery of the fractional quantum Hall effect and the advent 
and application of topological quantum field theories have revolutionized our understanding of the quantum properties of matter
\cite{Tsui_82,Laughlin_83,Arovas_84,PhysRevLett.49.405,PhysRevLett.95.146802,
PhysRevLett.98.106803,PhysRevLett.61.2015,Lundholm_2016}.
Among the prospective applications,
the notion of
topological quantum computation has recently emerged as one of the most exciting approaches for constructing a fault-tolerant quantum computer,
by seeking to exploit the emergent properties of many-particle systems to encode and manipulate quantum information in a manner which is resistant to error~\cite{K,lloyd2002quantum,freedman2003topological,Nayak_08}.
One simple such proposal for topological quantum computing 
and information storage
relies on the existence of topological states of matter whose quasiparticle excitations are anyons.

An anyon is a type of quasiparticle that can arise in 
systems confined to two dimensions and whose exchange properties
interpolate between bosons and fermions~\cite{leinaas1977theory,Wilczek_82,Wilczek_82b}. 
Because of the topological peculiarities of two spatial dimensions, the world lines 
of anyons can braid nontrivially around each other~\cite{leinaas1977theory,goldin1981representations,Wu_84}, 
and therefore, unlike fermions or bosons, exchanging two anyons twice is not topologically equivalent to leaving them alone. 
This opens up a whole new domain of quantum statistics known as intermediate or fractional statistics.
Even though the realization of anyons in experimentally feasible systems has been subject of recent research~\cite{CooSim-15,ZhaSreGemJai-14,
ZhaSreJai-15,MorTurPolWil-17,
UmuMacComCar-18,CorDubLunRou-19,
Yakaboylu_2018,Yakaboylu_anyon_2020}, all these works concern particles 
moving on the Euclidean plane $\mathbb{R}^2$, or a subset thereof. 
However, since the statistical behaviour of anyons depends on the topology, 
and even more importantly on the geometry and symmetry, of the underlying space, 
investigations on curved spaces can demonstrate novel properties of quantum 
statistics~\cite{Thouless_85,Einarsson_90,CMO,
einarsson1991fractional,
ouvry2019anyons,polychronakos2020two,
Haldane_83,Laughlin_83} (see also graph geometries \cite{Harrison-etal-14,MacSaw-19}). 
Indeed it has been recently demonstrated that the emerging fractional statistics for particles
restricted to move on the sphere, instead of on the plane, arises naturally in the context of quantum impurity problems, particularly, in the context of molecular impurities~\cite{Brooks_2021}. There, it has been shown that the emerging statistical interaction manifests itself in the alignment of molecules, which could also be of use as a powerful technique to measure the statistics parameter. This paves the way towards experimental realization as well as detection of anyons on the sphere using molecular impurities.

In the present manuscript, we explicitly show how the angulon Hamiltonian~\cite{Lemeshko_2015, PhysRevX.6.011012, Lemeshko_2016_book} gives rise to a system of two interacting anyons on the two-sphere $\mathbb{S}^2$. 
The angulon represents a quantum impurity exchanging orbital angular momentum with a many-particle bath, and serves as a reliable model for the rotation of molecules in superfluids~\cite{lemeshko2016quasiparticle, YuliaPhysics17, Shepperson16, Shepperson17}. 
In particular, we demonstrate that, under appropriate time-reversal symmetry breaking conditions, restricting the angulon Hamiltonian to states in the first Born-Oppenheimer approximation gives rise to the anyon Hamiltonian. 
Time-reversal symmetry is broken by using an additional external magnetic field and applying rotation, while the Born-Oppenheimer approximation is satisfied by considering heavy molecules. We further discuss and supply some technical details of the argument that had been 
left out in \cite{Brooks_2021}. Note that the phenomenon of quantum statistics 
transmutation typically involves emergent scalar interaction potentials 
and non-statistical gauge fields as well, and it is necessary to have sufficient
control of these effects in order to provide robust signatures of anyons.

\section{Anyon Hamiltonian}

Anyons are identical particles described by wave functions $\Psi$ which 
acquire a phase factor $e^{i\alpha\pi}$, respectively $e^{-i\alpha\pi}$, 
under permutation of two sets of coordinates. 
In contrast to fermions and bosons, we do not assume that the statistics parameter $\alpha$ is an integer. 
Namely, it could be any real number, say between 0 and 1, or between -1 and 0, 
i.e. a fraction of an integer (thereby `fractional statistics').
Consequently, we have to distinguish between the continuous exchange processes 
where two particles make an elementary anti-clockwise braid around each other, 
in which case the wave function gains a factor $e^{i\alpha\pi}$, 
and processes where they braid clockwise around each other, 
in which case the wave function has to acquire the 
inverse factor $e^{-i\alpha\pi}$. 
Here we however see a difference on the sphere compared to the plane,
since e.g. the 2-particle braid group reduces from $\mathbb{Z}$ on the plane
to $\mathbb{Z}_2$ on the sphere, due to a double exchange being topologically trivial.
This also means that we cannot 
determine topologically which 
way the particles braided,
and thus reduces the whole problem to the ordinary case of bosons or fermions, 
$\alpha \in \{0,1\}$. 
In fact, this conclusion is a manifestation of the symmetry of the full sphere,
and indeed the existence of anyons necessarily requires the breaking of 
time-reversal or orientation symmetry 
(corresponding to the choice of sign of $\alpha$ 
and the handedness of braids in our braid group representation).
A similar analysis for the $N$-particle case leads to the condition
$(N-1)\alpha \in \mathbb{Z}$, analogous to the well-known Dirac quantization condition 
\cite{Thouless_85,CMO,ouvry2019anyons}.
We can overcome this issue, by instead considering the punctured sphere 
$\mathbb{S}^2\setminus \{\mathcal{N}\}$, where $\mathcal{N}$ denotes the north pole (and $\mathcal{S}$ will denote the south pole). 
Clearly, $\mathbb{S}^2\setminus \{\mathcal{N}\}$ is topologically equivalent to the plane. 
Nevertheless, the analysis of anyons living on the sphere
(or a subset thereof)
requires 
novel ideas and techniques. The first reason for this is, that 
$\mathbb{S}^2\setminus \{\mathcal{N}\}$ carries a non-flat geometry, i.e. the 
free dynamics of two anyons is given by the Hamiltonian
\begin{align*}
H^{\mathrm{sing}}_{\mathrm{Anyon}} := -\sum_{j=1}^2\nabla_j^2:=-\sum_{j=1}^2 g_{ab}\nabla_j^a\nabla_j^b \, ,
\end{align*}
where $g_{ab}$ is the metric tensor of the sphere,
and we put suitable conditions at $\mathcal{N}$ 
and on the coincidence set for the particles (for 
simplicity, we may consider functions $\Psi$ vanishing on the diagonal of the
configuration space $(\mathbb{S}^2 \setminus \{\mathcal{N}\})^2$;
cf. \cite{BorSor-92,LunSol-14,CorOdd-18}.)
The second difference to the plane is, that the natural 
orientation-preserving symmetry group of the
full sphere is given by the three dimensional rotations $SO(3)$, while the symmetry 
group of the plane consists of a rotation around a single axis and translations 
in the plane. 
As one might expect, and we will see explicitly below,
the symmetry group plays a crucial role in deriving the emergence of anyons from suitable impurity problems.\\

It will be convenient to represent the anyonic wave function as 
$\Psi=e^{i\alpha\phi}\psi$, 
where $\psi$ is a bosonic wave function and $\phi$ is a fixed smooth multivalued 
function with the property
$\phi(q_2,q_1)=\phi(q_1,q_2)\pm \pi$ under simple continuous exchange of the two coordinates $q_j \in \mathbb{S}^2$,
in order that $\Psi$ acquires a correct phase factor $e^{\pm i\alpha \pi}$. A concrete example of such a function $\phi$ is given in complex stereographic coordinates $z_1,z_2\in \mathbb{C}$ by $\frac{1}{i}\log\left(\frac{z_1-z_2}{|z_1-z_2|}\right)$.
Applying the unitary transformation $e^{i\alpha\phi}$ to the free anyon dynamics yields
\begin{align*}
H_{\mathrm{Anyon}}:=e^{-i\alpha\phi}H^{\mathrm{sing}}_{\mathrm{Anyon}}e^{i\alpha\phi}=-\sum_{j=1}^2\left(\nabla_j+i\alpha\nabla_j \phi\right)^2=-\sum_{j=1}^2\left(\nabla_j+iA_j\right)^2 \, ,
\end{align*}
with the anyon statistics gauge field $A_j$ given by $A_j:=\alpha\nabla_j \phi$. 
Note that $H_{\mathrm{Anyon}}$, which is unitarily equivalent to the free anyon 
dynamics (although by a \emph{singular} gauge transformation, 
thereby changing the reference geometry)
has the advantage of acting 
on bosonic (single-valued) wave functions $\psi$.

\section{Emerging Gauge Field from the Angulon Hamiltonian}
\label{sec:Emerging Gauge Field from the Angulon Hamiltonian}
The angulon Hamiltonian for two rotors/impurities is defined by
\begin{align}
\label{eq:AngulonHamilton}
H_{\mathrm{angulon}}:=-\sum_{j=1}^2 \nabla_j^2+\sum_{l,m} \omega_l\ b_{l,m}^\dagger b_{l,m} + b^\dagger_{Z(q_1,q_2)}+b_{Z(q_1,q_2)} \, ,
\end{align}
where $-\nabla_j^2 = \mathbf{L}_j^2 = L_{jx}^2 + L_{jy}^2 + L_{jz}^2$ 
is the rotor Hamiltonian with variable $q_j \in \mathbb{S}^2$, 
$b_{l,m}$ are collective rotation modes of the bath, and
$b_{Z(q_1,q_2)}^{(\dagger)} = \sum_{l,m} Z_{l,m}(q_1,q_2)^{(*)} b_{l,m}^{(\dagger)}$ 
defines the coupling between these systems at the Fr\"{o}hlich level~\cite{Lemeshko_2015, PhysRevX.6.011012, Lemeshko_2016_book}. 
Note that this Hamiltonian is typically fully invariant under the action of $O(3)$, so that we cannot
expect any non-trivial anyons to emerge.

Instead, in the following, we aim to derive the statistics gauge field $A_j$
as emergent from the following \emph{modified} angulon Hamiltonian:
\begin{align}
\label{eq:ModifiedHamilton}
H'_{\mathrm{angulon},\Omega}:=-\sum_j \nabla_j^2+\Omega^2 V(q_1,q_2)+\sum_{l,m} \omega_l\ b_{l,m}^\dagger b_{l,m}+\Omega\Lambda_{\bar{q}}+b^\dagger_{Z(q_1,q_2)}+b_{Z(q_1,q_2)} \, ,
\end{align}
where $\bar{q}:=(q_1+q_2)/|q_1+q_2|$ is the normalized center of mass,
\begin{align*}
\Lambda_{\bar{q}}:=\sum_{l,m,m_1,m_2}m\ \overline{D^l_{m,m_2}(\alpha,\beta,\gamma)}D^l_{m,m_1}(\alpha,\beta,\gamma)b_{l,m_1}^\dagger b_{l,m_2}
\end{align*}
is the momentum operator aligned in the direction of $-\bar{q}$ which we define with the help of the Wigner matrix $D^l_{m,m_2}(\alpha,\beta,\gamma)$ where $\alpha,\beta,\gamma$ are the Euler angles of a rotation $R_{\alpha,\beta,\gamma}$ with the property $R_{\alpha,\beta,\gamma}(\mathcal{S})=\bar{q}$,
$V$ is an additional quadratic potential, and $\Omega$ is a large parameter. With the convention above, the momentum operator $\Lambda_z$ aligned with the $z-axis$ reads $\Lambda_z=\Lambda_\mathcal{S}$.
Note that having the momentum operator $\Lambda_{\bar{q}}$ aligned in the direction $\bar{q}$ will simplify our computation significantly. 
In the next section
we will discuss a model where we take
the operator $\Lambda_{z} 
= \Lambda_\mathcal{S}$ aligned with the $z$-axis as usual,
and argue that as $\Omega \to \infty$ they describe the same limit within a certain setup.
We will also discuss how one can realize the modified operator $H'_{\mathrm{angulon},\Omega}$, by coupling $H_{\mathrm{angulon}}$ to an additional constant magnetic field. 
In this concrete realization of Eq.~\eqref{eq:ModifiedHamilton} the scaling on $V$ comes naturally.

We refer to Hamiltonian (\ref{eq:ModifiedHamilton}) as modified, since it has a dispersion relation $\sum_{l,m} \omega_l\ b_{l,m}^\dagger b_{l,m}+\Omega\Lambda_{\bar{q}}$ which is not invariant under a change of orientation. 
Furthermore, the introduction of a suitably chosen potential $V$ 
punctures the sphere and therefore breaks the $SO(3)$ invariance as well. Let us denote with $\theta_j$ the azimuthal angle of $q_j$ and with $\varphi_j$ its polar angle w.r.t. the laboratory reference frame. 
The $q_1,q_2$-dependent coefficients of $Z(q_1,q_2)$ are then given by
\begin{align*}
Z_{l,m}:=\sum_{j}c_l\ Y_{l,m}(\theta_j,\varphi_j)\, ,
\end{align*}
where $Y_{l,m}$ are the spherical harmonics and $c_l$ are real coefficients. We will occasionally suppress the $q_1,q_2$-dependency of $Z(q_1,q_2)$, and simply write $Z$. 
Note that we may instead of Eq.~\eqref{eq:ModifiedHamilton} consider a
symmetry-breaking interaction such as $\tilde{Z}:=(1+\Omega\Lambda_{\bar{q}}\omega^{-1})^{-1}\cdot Z(q_1,q_2)$ leading to the emergence of anyons with the same statistical gauge field. As stated, however, we here aim for a simplest possible realization of anyons, as a first step.

The full Hamiltonian \eqref{eq:ModifiedHamilton} acts on an appropriate dense
domain in the tensor product Hilbert space of the impurities $L^2_{\mathrm{sym/asym}}(\mathbb{S}^2 \times \mathbb{S}^2)$, where $L^2_{\mathrm{sym}}(\mathbb{S}^2 \times \mathbb{S}^2)$ is the bosonic Hilbert space and $L^2_{\mathrm{asym}}(\mathbb{S}^2 \times \mathbb{S}^2)$ the fermionic one,
with the Fock space $\mathcal{F}\left(L^2\left(\mathbb{S}^2\right)\right)$ of the bath.
Following the analysis for impurity problems in the planar case \cite{Yakaboylu_anyon_2020},
the statistics gauge field emerges from $H'_{\mathrm{angulon},\Omega}$, by restricting it to the ground state of its pure many-body part 
$\sum_{l,m} \omega_l\ b_{l,m}^\dagger b_{l,m}+\Omega\Lambda_{\bar{q}}+b^\dagger_{Z(q_1,q_2)}+b_{Z(q_1,q_2)}$ which acts only on the Fock space $\mathcal{F}\left(L^2\left(\mathbb{S}^2\right)\right)$ of the bath.
Namely, with the help of a coherent state transformation, we can write the ground state as (we use the notation $b_x = \sum_{l,m} x_{lm} b_{lm}$ and $\cdot$ for action or composition)
\begin{align}
\label{eq:Vacuum}
\Phi(q_1,q_2):=\exp\left[b_{(\omega+\Omega\Lambda_{\bar{q}})^{-1}\cdot Z(q_1,q_2)}\ -\ b^\dagger_{(\omega+\Omega\Lambda_{\bar{q}})^{-1}\cdot Z(q_1,q_2)}\right]\cdot \ket{0}\, .
\end{align}
Explicitly, by completing the square,
$$
b^\dagger\cdot \left(\omega+\Omega\Lambda_{\bar{q}}\right)\cdot b+b^\dagger_Z+b_Z=\left(b+\xi\right)^\dagger\cdot \left(\omega+\Omega\Lambda_{\bar{q}}\right)\cdot \left(b+\xi\right)-c \, ,
$$
with $\xi:=(\omega+\Omega\Lambda_{\bar{q}})^{-1}\cdot Z$ and $c:=\xi^\dagger\cdot (\omega+\Omega\Lambda_{\bar{q}})\cdot \xi$. 
We see that the non-symmetric dispersion relation $\omega+\Omega\Lambda_{\bar{q}}$ leads to a breaking of symmetry in the vacuum section $\Phi(q_1,q_2)$, since the coefficients $(\omega+\Omega\Lambda_{\bar{q}})^{-1}\cdot Z(q_1,q_2)$ are no longer invariant under the action of $O(3)$.

In the following, we consider a gapped dispersion $\omega_l\rightarrow \infty$ and heavy impurities such that the ground state decouples from the rest of the Hamiltonian. In this regime, the low energy spectrum of Hamiltonian (\ref{eq:ModifiedHamilton}) can be described by the first Born-Oppenheimer approximation
\begin{align*}
\braket{\psi|H_{\mathrm{Emerg}}|\psi}:=\braket{\psi \Phi|H'_{\mathrm{angulon},\Omega}|\psi \Phi}\, ,
\end{align*}
where $\psi(q_1,q_2)$ is an impurity wave function (bosonic or fermionic). 
By applying the coherent state transformation $S_0:=\exp\left[b_{(\omega+\Omega\Lambda_{\bar{q}})^{-1}\cdot Z(q_1,q_2)}\ -\ b^\dagger_{(\omega+\Omega\Lambda_{\bar{q}})^{-1}\cdot Z(q_1,q_2)}\right]$ 
as above,
we see that, formally,
\begin{align*}
H_{\mathrm{Emerg}}=\braket{0|-\sum_j(\nabla_j+S_0^{-1}\left(\nabla_j S_0\right))^2|0}+\Omega^2 V-Z^\dagger(\omega+\Omega\Lambda_{\bar{q}})^{-1}Z \, .
\end{align*}
The issue with this representation is, that we do not have a nice expression for the quantity $S_0^{-1}\nabla_j S_0$. This is due to the fact that the following family of operators is non-commuting:
$$\{b_{(\omega+\Omega\Lambda_{\bar{q}})^{-1}\cdot Z(q_1,q_2)}\ -\ b^\dagger_{(\omega+\Omega\Lambda_{\bar{q}})^{-1}\cdot Z(q_1,q_2)}:q_1,q_2\in \mathbb{S}^2 \}\, ,$$ 
and therefore we cannot apply the usual chain rule $\nabla_j \exp[F]=\nabla_j F\exp[F]$. In order to arrive at an explicit expression, we will apply two unitary transformations, which should map the non-commuting family to a commuting one. Note that this transformation has to be $q_1,q_2$-dependent, since a single fixed unitary transformation always maps non-commuting families to non-commuting ones.

We first need to transform the whole system to a fixed reference point, such that 
$\bar{q} \mapsto \bar{q}'=\mathcal{S}$, i.e. such that the middlepoint between $q_1$ and $q_2$ stays fixed at the south pole $\mathcal{S}$. For an arbitrary point $q\neq \mathcal{N}$ which is not the north pole, let $T\in SO(3)$ be a rotation which maps the point $q$ into the south pole, i.e. $T(q)=\mathcal{S}$. Clearly there are many rotations, which satisfy $T(q)=\mathcal{S}$. Therefore, we demand further that $T$ leaves the axis $\mathcal{S}\times q$ invariant for $q\neq \mathcal{S}$ 
and define $T$ to be the identity if $q=\mathcal{S}$.
The conditions $T(q)=\mathcal{S}$ and $T(\mathcal{S}\times q)=\mathcal{S}\times q$ uniquely determine the map $T$. Since $T$ is $q$-dependent, we will write $T_q=T$. In the following, we will always use the middle point $\bar{q}$ as the argument, i.e we consider $T_{\bar{q}}$. In order to promote $T_{\bar{q}}$ to a transformation on the whole Hilbert space, note that we can write it as 
\begin{align*}
T_{\bar{q}}=\exp\left[\left(\begin{array}{rrr}
0 & -z_{\bar{q}} & y_{\bar{q}}\\
z_{\bar{q}} & 0 & -x_{\bar{q}}\\
-y_{\bar{q}} & x_{\bar{q}} & 0
\end{array}\right)\right]\, ,
\end{align*}
with coefficients $(x_{\bar{q}},y_{\bar{q}},z_{\bar{q}}):=d(\bar{q},\mathcal{S}) (\mathcal{S}\times \bar{q})/|\mathcal{S}\times \bar{q}|$, where $d(\bar{q},\mathcal{S})$ is the geodesic distance of $\bar{q}$ to the south pole $\mathcal{S}$. Let us furthermore denote transformed points as $q':=T_{\bar{q}}\cdot q$. With this at hand, we can define the transformation of a Fock space valued state $\Psi(q_1,q_2)$ as
\begin{align*}
\hat{T}\left(\Psi\right)(q_1,q_2):=\exp\left[i\left(x_{\bar{q}} \Lambda_x+y_{\bar{q}}\Lambda_y+z_{\bar{q}}\Lambda_z\right)\right]\cdot \Psi\left(q_1,q_2\right)\, .
\end{align*}
Recall, that the transformation $T_{\bar{q}}$ only makes sense as long as $\bar{q}\neq \mathcal{N}$. Therefore, we only consider this transformation $\hat{T}$ 
for confined states $\Psi$, for example only for states which have a support contained in an open set $O\subset \overline{O}\subset \{q\in \mathbb{S}^2:q_3<0\}$. 
Note that this is not necessarily a real restriction, since the modified operator $H'_{\mathrm{angulon},\Omega}$ contains a confining potential $V$ anyway, which we will assume to have its minimum close to $\mathcal{S}$.\\

We can write the transformed Hamiltonian $\hat{T}^{-1}H'_{\mathrm{angulon},\Omega}\hat{T}$ as
\begin{align*}
-\sum_j \left(\nabla_j+\hat{T}^{-1}\left(\nabla_j \hat{T}\right)\right)^2+\Omega^2 V(q_1,q_2)+\sum_{l,m} \omega_l\ b_{l,m}^\dagger b_{l,m}+\Omega\Lambda_{z}+b^\dagger_{Z(q'_1,q'_2)}+b_{Z(q'_1,q'_2)}\, .
\end{align*}
Note that after the transformation, the angular momentum operator $\Lambda_{z}$ is aligned with respect to the $z$-axis instead of the direction $\bar{q}$. Let us denote with $\phi=\phi(q_1,q_2)$ the polar angle of the transformed point $q'_1$. Furthermore, let $R=R_{\phi}$ be a rotation around the $z$-axis by an amount of $\phi$. Then, the polar angle of the transformed point $q_1'':=R^{-1}(q_1')$ is zero, while the polar angle of $q_2'':=R^{-1}(q_2')$ equals $\pi$. Both points $q_1'',q_2''$ have the same azimuthal angle $\theta$. 
We promote $R=R_\phi$ to an operation on the whole Hilbert space by
\begin{align*}
\hat{R}\left(\Psi\right)(q_1,q_2):=\exp\left[i\phi\Lambda_z\right]\cdot \Psi\left(q_1,q_2\right) \, .
\end{align*}
The transformed operator $\hat{R}^{-1}\hat{T}^{-1}H'_{\mathrm{angulon},\Omega}\hat{T}\hat{R}$ then reads
\bals
-&\sum_j \left(\nabla_j+\hat{R}^{-1}\hat{T}^{-1}\left(\nabla_j \hat{T}\right)\hat{R}+i\left(\nabla_j \phi\right)\sum_{l,m}m\ b_{l,m}^\dagger b_{l,m}\right)^2+\Omega^2 V(q_1,q_2)+\sum_{l,m} \omega_l\ b_{l,m}^\dagger b_{l,m}\\ 
&+\Omega\Lambda_{z}+b^\dagger_{Z(q''_1,q''_2)}+b_{Z(q''_1,q''_2)} \, ,
\eals
In the final step, we diagonalize the pure many body part $\sum_{l,m} \omega_l\ b_{l,m}^\dagger b_{l,m}+\Omega\Lambda_{z}+b^\dagger_{Z(q''_1,q''_2)}+b_{Z(q''_1,q''_2)}$, by applying the coherent state transformation
\begin{align*}
S:=\exp\left[b_{(\omega+\Omega\Lambda_z)^{-1}\cdot Z(q_1'',q_2'')}\ -\ b^\dagger_{(\omega+\Omega\Lambda_z)^{-1}\cdot Z(q_1'',q_2'')}\right]\, .
\end{align*}
Since the coefficients at the transformed points $Z_{l,m}(q_1'',q_2'')=(1+(-1)^m)c_l Y_{l,m}(\theta,0)$ are all real valued and the expressions only dependent on $\theta$,
we know that the collection 
$$\Bigl\{ b_{(\omega+\Omega\Lambda_z)^{-1}\cdot Z(q_1'',q_2'')}\ -\ b^\dagger_{(\omega+\Omega\Lambda_z)^{-1}\cdot Z(q_1'',q_2'')}:q_1,q_2\in \mathbb{S}^2\setminus \{O\}  \Bigr\}$$ 
is a family of commuting operators. Consequently, we can finally apply the chain rule and compute $S^{-1}\left(\nabla_j S\right)$ quite explicitly as
\begin{align*}
S^{-1}\left(\nabla_j S\right)=b_{(\omega+\Omega\Lambda_z)^{-1}\cdot \nabla_j Z(q_1'',q_2'')}\ -\ b^\dagger_{(\omega+\Omega\Lambda_z)^{-1}\cdot \nabla_j Z(q_1'',q_2'')}\, .
\end{align*}
We can express the transformed Hamiltonian $S^{-1}\hat{R}^{-1}\hat{T}^{-1}H'_{\mathrm{angulon},\Omega}\hat{T}\hat{R}S$ as
\begin{align*}
-\sum_j&\left(\nabla_j+i\alpha(\theta)\left(\nabla_j \phi\right)+Y_j+S^{-1}\left(\nabla_j S\right)-i\left(\nabla_j\phi\right)\left(b_{W}+b^\dagger_{W}\right)+i\left(\nabla_j\phi\right)\sum_{l,m}m\ b_{l,m}^\dagger b_{l,m}\right)^2\\
& \ \ \ \ +\Omega^2 V(q_1,q_2)-E_0+\sum_{l,m} \omega_l\ b_{l,m}^\dagger b_{l,m}+\Omega\Lambda_{z}\, ,
\end{align*}
with the abbreviations
$$W:=\Lambda_z(\omega+\Omega\Lambda_z)^{-1}Z(q_1'',q_2'')\, ,$$ 
$$Y_j:=S^{-1}\hat{R}^{-1}\hat{T}^{-1}\left(\nabla_j \hat{T}\right)\hat{R} S\, ,$$ 
$$E_0:=Z(q_1'',q_2'')^T(\omega+\Omega\Lambda_z)^{-1}Z(q_1'',q_2'')\, ,$$ and
\begin{align*}
\alpha(\theta):=Z(q_1'',q_2'')^T(\omega+\Omega\Lambda_z)^{-1}\Lambda_z(\omega+\Omega\Lambda_z)^{-1}Z(q_1'',q_2'')\, .
\end{align*}
Observe that the vacuum expectation of $i\nabla_j\phi\left(b_{W}+b^\dagger_{W}\right)$ and $i\nabla_j\phi\sum_{l,m}m\ b_{l,m}^\dagger b_{l,m}$ is zero. Therefore these terms will only contribute to the emergent scalar potential but not to the 
emergent gauge field. Let us recall the definition of the vacuum section $\Phi$ in equation (\ref{eq:Vacuum}). With the help of the unitary maps $T,R$ and $S$ we can write $\Phi=\hat{T}\hat{R}S\ket{0}$ and consequently
\bals
 H_{\mathrm{Emerg}}& :=\braket{\Phi|H'_{\mathrm{angulon},\Omega}|\Phi}=\braket{0|S^{-1}\hat{R}^{-1}\hat{T}^{-1}H'_{\mathrm{angulon},\Omega}\hat{T}\hat{R}S|0} \\
&=- \bra{0}\sum_j\left(\nabla_j+i\alpha(\theta)\left(\nabla_j \phi\right)+Y_j+S^{-1}\left(\nabla_j S\right)-i\left(\nabla_j\phi\right)\left(b_{W}+b^\dagger_{W}\right) \right. \\
& \left. + i\left(\nabla_j\phi\right)\sum_{l,m}m\ b_{l,m}^\dagger b_{l,m}\right)^2\ket{0} +\braket{0|\Omega^2 V-E_0+\sum_{l,m} \omega_l\ b_{l,m}^\dagger b_{l,m}+\Omega\Lambda_{z}|0}\, .
\eals
The second expectation is simply $\braket{0|\Omega^2 V-E_0+\sum_{l,m} \omega_l\ b_{l,m}^\dagger b_{l,m}+\Omega\Lambda_{z}|0}=\Omega^2V-E_0$. In order to compute the first one, let us define the magnetic potential $A_j:=\alpha(\theta)\nabla_j \phi$ and the gauge covariant derivative $\nabla_j^A:=\nabla_j+iA_j$. Let us further abbreviate 
\begin{align*}
U_j:=(\omega+\Omega\Lambda_z)^{-1}\left[\nabla_j Z(q_1'',q_2'')-i\nabla_j\phi\Lambda_z Z(q_1'',q_2'')\right]\, .
\end{align*}
Then we can rewrite the first vacuum expectation as
\begin{align*}
&\braket{0|-\sum_j\left(\nabla^A_j+Y_j+\left(b_{U_j}+b^\dagger_{U_j}\right)+i\left(\nabla_j\phi\right)\sum_{l,m}m\ b_{l,m}^\dagger b_{l,m}\right)^2|0}\\
&\ \ =-\sum_j\Big[\left(\nabla_j^A+\braket{0|Y_j|0}\right)^2-\braket{0|Y_j|0}^2+\braket{0|(Y_j+b_{U_j}+b^\dagger_{U_j}+i\left(\nabla_j\phi\right) \sum_{l,m}m\ b_{l,m}^\dagger b_{l,m})^2|0}\Big]\, .
\end{align*}
Let us define the magnetic background potential $\tilde{A}_j:=-i\braket{0|Y_j|0}$ and the modified scalar potential 
\begin{align*}
\widetilde{V}:=\Omega^2V+\sum_j \braket{0|Y_j|0}^2-\sum_j\braket{0|(Y_j+b_{U_j}+b^\dagger_{U_j}+i\left(\nabla_j\phi\right) \sum_{l,m}m\ b_{l,m}^\dagger b_{l,m})^2|0}-E_0\, ,
\end{align*}
then we can compactly express the emergent Hamiltonian as
\begin{align*}
H_{\mathrm{Emerg}}=-\sum_j \left(\nabla_j^A+i\tilde{A}_j\right)^2+\widetilde{V}\, .
\end{align*}

In case of constant $\alpha(\theta)=\alpha$, the operator $-\sum_j \left(\nabla_j^A\right)^2$ corresponds to the anyon Hamiltonian $H_{\mathrm{Anyon}}$ with statistics parameter $\alpha$, i.e. 
\begin{align}
\label{eq:EmergingAnyon}
H_{\mathrm{Emerg}}=H_{\mathrm{Anyon}}+\sum_j \left(\tilde{A}_j^2-i\left(\nabla_j\cdot \tilde{A}_j\right)-2i\tilde{A}_j\cdot \nabla_j^A\right)+\widetilde{V}\, .
\end{align}
In Ref.~\cite{Brooks_2021}, approximately constant $\alpha$ (depending on $\Omega$)
is indeed realized for a suitable, and experimentally feasible, choice of $\omega$ and $c_l$. 
Particularly, $\omega$ is chosen at the roton minimum of the dispersion relation of superfluid helium, which allows us to achieve a gapped dispersion, and the coupling $c_l$ is described by the model interaction used in order to describe angulon instabilities and oscillations observed in the experiment.
Therefore, the Hamiltonian $H'_{\mathrm{angulon},\Omega}$ gives rise to a system of two anyons, coupled to an additional magnetic potential $\tilde{A}_j$ and an additional scalar potential $\widetilde{V}$.\\

In the following let us verify that the magnetic potential $\tilde{A}_j$ is regular, 
which on a suitable scale means that $\mathrm{curl}\,\tilde{A}_j$ can be treated as a background field 
and thus does not influence the statistics. First of all, we can write it as
\begin{align*}
\tilde{A}_j&=-i\braket{0|S^\dagger\hat{R}^\dagger\hat{T}^{-1}\nabla_j(\hat{T})\hat{R}S|0}=-i\braket{\Phi|\hat{T}^{-1}\nabla_j(\hat{T})|\Phi}\, ,
\end{align*}
where $\Phi$ is as usual the vacuum section. First we want to compute the Lie algebra element $T_{\bar{q}}^{-1}\left(\nabla_j T_{\bar{q}}\right)$. In order to verify that $T_{\bar{q}}$ is a matrix-valued $C^\infty$ (smooth) function, so especially that its derivative exists and is a continuous function, recall the explicit representation
\begin{align*}
T_{\bar{q}}=\exp\left[\frac{d(\bar{q},\mathcal{S})}{|\mathcal{S}\times \bar{q}|}\left(\begin{array}{rrr}
0 & -\left(\mathcal{S}\times \bar{q}\right)_3 & \left(\mathcal{S}\times \bar{q}\right)_2\\
\left(\mathcal{S}\times \bar{q}\right)_3 & 0 & -\left(\mathcal{S}\times \bar{q}\right)_1\\
-\left(\mathcal{S}\times \bar{q}\right)_2 & \left(\mathcal{S}\times \bar{q}\right)_1 & 0
\end{array}\right)\right]\, .
\end{align*}
As long as $\mathcal{S}\times \bar{q}\neq 0$, i.e. as long as $\bar{q}\neq \mathcal{S}$ and $\bar{q}\neq \mathcal{N}$, $T_{\bar{q}}$ is clearly $C^\infty$. Since we want to investigate the limit $q_1,q_2\rightarrow \mathcal{S}$ anyway, we do not have to worry about the case $\bar{q}= \mathcal{N}$. Regarding the south pole itself, observe that the function $d(\bar{q},\mathcal{S})/|\mathcal{S}\times \bar{q}|$ is $C^\infty$, even for $\bar{q}=\mathcal{S}$. Consequently, we know that $T_{\bar{q}}^{-1}\left(\nabla_jT_{\bar{q}}\right)$ exists and it is a smooth function as long as $\bar{q}\neq \mathcal{N}$. Note that $T_{\bar{q}}^{-1}\left(\nabla_jT_{\bar{q}}\right)$ is an element of the Lie algebra of $SO(3)$, and therefore we can write it as
\begin{align*}
T_{\bar{q}}^{-1}\left(\nabla_jT_{\bar{q}}\right)=\left(\begin{array}{rrr}
0 & -c(q_1,q_2) & b(q_1,q_2)\\
c(q_1,q_2) & 0 & -a(q_1,q_2)\\
-b(q_1,q_2) & a(q_1,q_2) & 0
\end{array}\right)\, ,
\end{align*}
with continuous and real functions $a(q_1,q_2),b(q_1,q_2)$ and $c(q_1,q_2)$. Consequently, we can write the operator $\hat{T}^{-1}\left(\nabla_j\hat{T}\right)$ as
\begin{align*}
\hat{T}^{-1}\left(\nabla_j\hat{T}\right)= a_j(q_1,q_2) i\Lambda_x+ b_j(q_1,q_2) i\Lambda_y+ c_j(q_1,q_2) i\Lambda_z \, .
\end{align*}
From the representation above we see that the additional magnetic field $\tilde{A}_j$ is regular and therefore does not contribute to the statistics, i.e. $H_{\mathrm{Emerg}}$ describes anyons subject to an additional magnetic gauge
field $\tilde{A}_j(q_1,q_2)$ as well as an additional electric potential field $\widetilde{V}(q_1,q_2)$. Let us now describe a set-up, where the additional magnetic background field $\tilde{A}=(\tilde{A}_1,\tilde{A}_2)$ can be neglected entirely, i.e. we look for reasonable conditions such that $\tilde{A}\underset{q_1,q_2\rightarrow \mathcal{S}}{\longrightarrow}0$. 
Since $T$ and $\Phi$ are compatible with rotations around the $z$ axis, we know that $\tilde{A}(q_1=\mathcal{S},q_2=\mathcal{S})=0$. It is therefore enough to verify that the vector field $\tilde{A}$ is continuous. 
By our representation of $\hat{T}^{-1}\left(\nabla_j\hat{T}\right)$ above, the continuity of $\tilde{A}$ follows from the continuity of $\braket{\Phi(q_1,q_2)| \Lambda_e |\Phi(q_1,q_2)}$, $e\in \mathbb{S}^2$, 
in $q_1,q_2$. A sufficient condition for $\braket{\Phi(q_1,q_2)| \Lambda_e |\Phi(q_1,q_2)}$ being continuous would be the following growth condition on the coefficients: 
$|c_l|\leq \frac{C}{l}$ and $\omega_l\pm\Omega l\geq l^{1+\epsilon}$ with $\epsilon>0$.
While the former condition can be fulfilled with the model parameter describing the molecule-helium interaction, see Ref.~\cite{Brooks_2021}, the latter condition can be satisfied by considering a strong magnetic field. 
Then, $\tilde{A}\underset{q_1,q_2\rightarrow \mathcal{S}}{\longrightarrow}0$ vanishes for $q_1,q_2$ close to the south pole, in contrast to the singular gauge field $\nabla_j \phi$ which has a pole at $\mathcal{S}$.

With the convergence $\tilde{A}\underset{q_1,q_2\rightarrow \mathcal{S}}{\longrightarrow}0$ at hand, we can verify that the coupling to the background field $\nabla_j^A\mapsto \nabla_j^A+\tilde{A}_j$ can be neglected in the limit of large $\Omega$. In order to do this, let us define the dilatation operator
\begin{align*}
D_\ell:
\begin{cases}
&L^2\left(\ell \mathbb{S}^2\right)\otimes L^2\left(\ell \mathbb{S}^2\right)\rightarrow L^2\left(\mathbb{S}^2\right)\otimes L^2\left(\mathbb{S}^2\right),\\
&\Psi\mapsto D_\ell\left(\Psi\right)(q_1,q_2):=\Psi(\ell q_1,\ell q_2)\, .
\end{cases}.
\end{align*}
Note that the statistics gauge field $A$ transforms exactly like the derivative operator as $D_\ell^{-1} A D_\ell=\ell A$. Therefore, $D_\ell^{-1}\nabla_j^A D_\ell=\ell\nabla_j^A$ and $D_\ell^{-1}H_{\mathrm{Anyon}}D_\ell=\ell^2 H_{\mathrm{Anyon}}$. Transforming the emerging Hamiltonian yields
\begin{align*}
D_\ell^{-1}\left(\ell^{-2}H_{\mathrm{Emerg}}\right)D_\ell&=-\sum_j \left(\nabla_j^A+i\ell^{-1}\tilde{A}\left(\ell^{-1}q_1,\ell^{-1}q_2\right)\right)^2 +\ell^{-2}\widetilde{V}(\ell^{-1}q_1,\ell^{-1}q_2)\, .
\end{align*}
Using the assumption
that the confining potential $V$ is quadratic, we see that the natural length scale of the confinement is given by the equation $\ell^{-4}\Omega^2$, i.e. $\ell:=\sqrt{\Omega}$. In the next section we will see that a quadratic potential comes naturally, however different choices for $V$ could yield other interesting limits. Since $\ell\rightarrow \infty$ in the limit of large $\Omega$, we conclude $\tilde{A}\left(\ell^{-1}q_1,\ell^{-1}q_2\right)\rightarrow 0$. With the abbreviation $V_\Omega(q_1,q_2):=\frac{1}{\sqrt{\Omega}}\widetilde{V}(\frac{q_1}{\sqrt{\Omega}},\frac{q_2}{\sqrt{\Omega}})$ 
we then have the asymptotic result
\begin{align*}
D^{-1}_{\sqrt{\Omega}}\left(\frac{1}{\sqrt{\Omega}} H_{\mathrm{Emerg}}\right)D_{\sqrt{\Omega}}\underset{\Omega\rightarrow \infty}{\longrightarrow} \sqrt{\Omega}\ H_{\mathrm{Anyon}}+V_\Omega\, .
\end{align*}
Hence, in the limit of large $\Omega$, the emerging Hamiltonian corresponds to a system of two anyons living on a sphere of radius $\ell=\sqrt{\Omega}$, 
with no additional magnetic field but coupled only to an additional scalar potential $V_\Omega$.

\section{Realization of a modified Quantum Dispersion Relation}
\label{sec:Realization of a modified Quantum Dispersion Relation}
We now come back to the standard angulon Hamiltonian 
\eqref{eq:AngulonHamilton}, and modify it by coupling it to an external
magnetic potential $A_j^\Omega:=\Omega(-y_j,x_j,0)$, i.e. we consider the operator
\begin{align}
\label{eq:RealizedHamilton}
H_{\mathrm{angulon},\Omega}:=-\sum_j \left(\nabla_j-iA_j^\Omega\right)^2+\sum_{l,m} \omega_l\ b_{l,m}^\dagger b_{l,m}+b^\dagger_Z+b_Z \, .
\end{align}
Let $J_z=L_{z,1}+L_{z,2}+\Lambda_z$ be the total angular momentum of the two particles together with the many body environment. By rotating the system in the $x-y$ plane at the cyclotron frequency $\Omega$ we obtain
\begin{align*}
\tilde{H}_{\mathrm{angulon},\Omega}:&=e^{-it\Omega J_z}\left(H_{\mathrm{angulon},\Omega}-i\partial_t\right)e^{it\Omega J_z}+i\partial_t\\
&=-\sum_j \left(\left(\nabla_j-iA_j^\Omega\right)^2+\Omega L_{z,j}\right)+\sum_{l,m} \omega_l\ b_{l,m}^\dagger b_{l,m}+\Omega \Lambda_z+b^\dagger_Z+b_Z\\
&=-\sum_j \nabla_j^2+\Omega^2\sum_j(x_j^2+y_j^2)+\sum_{l,m} \omega_l\ b_{l,m}^\dagger b_{l,m}+\Omega\Lambda_z+b^\dagger_Z+b_Z \, ,
\end{align*}
where we used $\Omega L_{z,j}=-iA^\Omega_{j}\cdot \nabla_j$. With the definition $V(q_1,q_2):=\sum_j(x_j^2+y_j^2)$, we see that $\tilde{H}_{\mathrm{angulon},\Omega}$ almost coincides with 
\begin{align*}
H'_{\mathrm{angulon},\Omega}=-\sum_j \nabla_j^2+\Omega^2 V(q_1,q_2)+\sum_{l,m} \omega_l\ b_{l,m}^\dagger b_{l,m}+\Omega\Lambda_{\bar{q}}+b^\dagger_Z+b_Z \, ,
\end{align*}
except that the angular momentum operator $\Lambda_{z}$ in $\tilde{H}_{\mathrm{angulon},\Omega}$ is aligned in the $z$ direction, while for $H'_{\mathrm{angulon},\Omega}$ the operator $\Lambda_{\bar{q}}$ is aligned in the center of mass direction $\bar{q}$.\\

In the previous section we have seen that the modified angulon Hamiltonian $H'_{\mathrm{angulon},\Omega}$ gives rise to a system of two interacting anyons in the limit of large $\Omega$. In the following, we want to argue why the same conclusion
holds for the slightly different operator $\tilde{H}_{\mathrm{angulon},\Omega}$, i.e.
we are going to justify that anyons emerge in the low energy regime of $\tilde{H}_{\mathrm{angulon},\Omega}$ as well. Let us define the Fock space valued function
\begin{align*}
\Psi:=\exp\left[b_{(\omega+\Omega\Lambda_z)^{-1}\cdot Z}\ -\ b^\dagger_{(\omega+\Omega\Lambda_z)^{-1}\cdot Z}\right]\cdot \ket{0} \, ,
\end{align*}
which is the vacuum state of the many-body part of $\tilde{H}_{\mathrm{angulon},\Omega}$, i.e. it is the ground state of
\begin{align*}
\mathbb{H}_\Omega:=\sum_{l,m} \omega_l\ b_{l,m}^\dagger b_{l,m}+\Omega\Lambda_z+b^\dagger_Z+b_Z \, .
\end{align*}
In order to observe the emergence of anyons, let us define the alternative section
\begin{align*}
\Phi:=\exp\left[b_{(\omega+\Omega\Lambda_{\bar{q}})^{-1}\cdot Z}\ -\ b^\dagger_{(\omega+\Omega\Lambda_{\bar{q}})^{-1}\cdot Z}\right]\cdot \ket{0} \, ,
\end{align*}
which gives rise to the correct gauge field $A_z$. The issue is, that $\Phi$ is no longer the vacuum section of $\mathbb{H}_\Omega$.
However, if we can show that $E_\Phi:=\braket{\Phi|\mathbb{H}_\Omega|\Phi}$ approximates the true ground state energy of $\mathbb{H}_\Omega$
\begin{align*}
\tilde{E}_0:=-Z^\dagger(\omega+\Omega \Lambda_z)^{-1}Z \, ,
\end{align*}
and if there is a spectral gap from a nondegenerate ground state, 
then we can argue that the states (considered as rays of the Hilbert space) are close.

Let us now verify that the energy deviation $\epsilon:=E_\Phi-\tilde{E}_0$ is small in the limit of large $\Omega$. First of all, we can express $\epsilon$ as
\begin{align*}
\epsilon=Z^\dagger\left[(\omega+\Omega\Lambda_{\bar{q}})^{-1}\Omega(\Lambda_z-\Lambda_{\bar{q}})(\omega+\Omega\Lambda_{\bar{q}})^{-1}-(\omega+\Omega\Lambda_{z})^{-1}\Omega(\Lambda_z-\Lambda_{\bar{q}})(\omega+\Omega\Lambda_{\bar{q}})^{-1}\right]Z \, .
\end{align*}
Let us make the reasonable assumption, that $\omega+\Omega\Lambda_{\bar{q}}$  is non degenerate, i.e. 
for simplicity let us assume that $\omega_l\pm\Omega l\geq \delta \Omega (l+1)$
for some $0<\delta<1$. Furthermore, we assume that $|Z_{l,m}|\leq \frac{C}{l+1}$ for some $C$. We define the operator 
$\nu$ by $\nu_l:=\sqrt{\delta (l+1)}$ in the diagonalizing basis of $\Lambda_z$. Since $\Lambda_{\bar{q}}$ is block-diagonal with respect to $l$, $\Lambda_{\bar{q}}$ commutes with $\nu$ and consequently we can rewrite the first part of the error $\epsilon=\epsilon_1-\epsilon_2$ (the second part can be rewritten in the same way)
\begin{align*}
\epsilon_1=(\nu^{-1}Z)^\dagger\left[\Omega\nu^2(\omega+\Omega\Lambda_{\bar{q}})^{-1}\Omega^{-1}\nu^{-1}(\Lambda_z-\Lambda_{\bar{q}})\nu^{-1}(\omega+\Omega\Lambda_{\bar{q}})^{-1}\nu^2\right](\nu^{-1}Z) \, .
\end{align*}
Note that we have the bound on the operator norm $\|\Omega\nu^2(\omega+\Omega\Lambda_{\bar{q}})^{-1}\|\leq 1$, as well as $\|\nu^{-1}(\Lambda_z-\Lambda_{\bar{q}})\nu^{-1}\|\leq \frac{2}{\delta}$ and $\|\chi^{< L}\nu^{-1}(\Lambda_z-\Lambda_{\bar{q}})\nu^{-1}\| \ll 1$
for $\bar{q}\rightarrow \mathcal{S}$, where we define $\chi^{< L}_l:=1$ for $l< L$ and $\chi^{< L}_l:=0$ otherwise. By our assumption
$|Z_{l,m}|\leq \frac{C}{l}$, we know that $\|\nu^{-1}Z\|^2=:\widetilde{C}<\infty$, therefore we obtain $\|(1-\chi^{<L}_\ell)\nu^{-1}Z\|\underset{L\rightarrow \infty}{\longrightarrow}0$ and
\bals
& \lim_{\bar{q}\rightarrow \mathcal{S}}\Omega \epsilon_1\leq \lim_{\bar{q}\rightarrow \mathcal{S}}2\left(\widetilde{C}\|\chi^{< L}\nu^{-1}(\Lambda_z-\Lambda_{\bar{q}})\nu^{-1}\|+\frac{2}{\delta}\|(1-\chi^{<L}_\ell)\nu^{-1}Z\|^2\right) \\
& =\frac{4}{\delta}\|(1-\chi^{<L}_\ell)\nu^{-1}Z\|^2\underset{L\rightarrow \infty}{\longrightarrow}0 \, .
\eals
Applying a similar argument for $\epsilon_2$ yields the estimate for the total error $\epsilon\ll \frac{1}{\Omega}$. Note that the ground state energy itself satisfies $\tilde{E}_0\approx \frac{1}{\Omega}$. Consequently, the error term $\epsilon$ is negligibly small even compared to the ground state energy $\tilde{E}_0$, i.e.
\begin{align*}
\epsilon \ll \tilde{E}_0 \, .
\end{align*}
We conclude that $E_\Phi:=\braket{\Phi|\mathbb{H}_\Omega|\Phi}$ approximates the ground state energy $\tilde{E}_0=\braket{\Psi|\mathbb{H}_\Omega|\Psi}$ of $\mathbb{H}_\Omega$, and since $\mathbb{H}_\Omega$ has a uniform spectral gap this especially means that there exists $\theta_{\bar{q}}\in [-\pi,\pi)$ such that $\|\Phi-e^{i\theta_{\bar{q}}}\Psi\|\ll 1$. This justifies the usage of the section $\Phi$ instead of $\Psi$ in the Born-Oppenheimer approximation.

Lastly, we stress that the Born-Oppenheimer approximation itself 
and the emergence of the exact anyonic spectrum was justified both analytically and numerically
for simpler but highly representative models in \cite{Yakaboylu_anyon_2020,Brooks_2021}.

\section{Conclusions}

In conclusion, we explicitly show that in the Born-Oppenheimer approximation the many-particle bath manifests itself as the statistics gauge field on the two-sphere with respect to the molecular impurities immersed into it. We further demonstrate that a possible experimental realization is feasible within the framework of the angulon quasiparticle by applying an external magnetic field to the molecular impurities and rotating the impurity-bath system. This lays the foundations for realizing anyons on the two-sphere in terms of molecular impurities in superfluid helium. We finally note that the dispersion relation of superfluid helium nanodroplets at the roton minumum allows us to explore the problem with a gapped dispersion relation so that the Born-Oppenheimer approximation can be achieved by considering heavy impurities as discussed in Ref.~\cite{Brooks_2021}.

\bibliography{AnyonsBib}

\begin{thebibliography}{45}%
\makeatletter
\providecommand \@ifxundefined [1]{%
 \@ifx{#1\undefined}
}%
\providecommand \@ifnum [1]{%
 \ifnum #1\expandafter \@firstoftwo
 \else \expandafter \@secondoftwo
 \fi
}%
\providecommand \@ifx [1]{%
 \ifx #1\expandafter \@firstoftwo
 \else \expandafter \@secondoftwo
 \fi
}%
\providecommand \natexlab [1]{#1}%
\providecommand \enquote  [1]{``#1''}%
\providecommand \bibnamefont  [1]{#1}%
\providecommand \bibfnamefont [1]{#1}%
\providecommand \citenamefont [1]{#1}%
\providecommand \href@noop [0]{\@secondoftwo}%
\providecommand \href [0]{\begingroup \@sanitize@url \@href}%
\providecommand \@href[1]{\@@startlink{#1}\@@href}%
\providecommand \@@href[1]{\endgroup#1\@@endlink}%
\providecommand \@sanitize@url [0]{\catcode `\\12\catcode `\$12\catcode
  `\&12\catcode `\#12\catcode `\^12\catcode `\_12\catcode `\%12\relax}%
\providecommand \@@startlink[1]{}%
\providecommand \@@endlink[0]{}%
\providecommand \url  [0]{\begingroup\@sanitize@url \@url }%
\providecommand \@url [1]{\endgroup\@href {#1}{\urlprefix }}%
\providecommand \urlprefix  [0]{URL }%
\providecommand \Eprint [0]{\href }%
\providecommand \doibase [0]{http://dx.doi.org/}%
\providecommand \selectlanguage [0]{\@gobble}%
\providecommand \bibinfo  [0]{\@secondoftwo}%
\providecommand \bibfield  [0]{\@secondoftwo}%
\providecommand \translation [1]{[#1]}%
\providecommand \BibitemOpen [0]{}%
\providecommand \bibitemStop [0]{}%
\providecommand \bibitemNoStop [0]{.\EOS\space}%
\providecommand \EOS [0]{\spacefactor3000\relax}%
\providecommand \BibitemShut  [1]{\csname bibitem#1\endcsname}%
\let\auto@bib@innerbib\@empty
\bibitem [{\citenamefont {Tsui}\ \emph {et~al.}(1982)\citenamefont {Tsui},
  \citenamefont {Stormer},\ and\ \citenamefont {Gossard}}]{Tsui_82}%
  \BibitemOpen
  \bibfield  {author} {\bibinfo {author} {\bibfnamefont {D.~C.}\ \bibnamefont
  {Tsui}}, \bibinfo {author} {\bibfnamefont {H.~L.}\ \bibnamefont {Stormer}}, \
  and\ \bibinfo {author} {\bibfnamefont {A.~C.}\ \bibnamefont {Gossard}},\
  }\href {\doibase 10.1103/PhysRevLett.48.1559} {\bibfield  {journal} {\bibinfo
   {journal} {Phys. Rev. Lett.}\ }\textbf {\bibinfo {volume} {48}},\ \bibinfo
  {pages} {1559} (\bibinfo {year} {1982})}\BibitemShut {NoStop}%
\bibitem [{\citenamefont {Laughlin}(1983)}]{Laughlin_83}%
  \BibitemOpen
  \bibfield  {author} {\bibinfo {author} {\bibfnamefont {R.~B.}\ \bibnamefont
  {Laughlin}},\ }\href {\doibase 10.1103/PhysRevLett.50.1395} {\bibfield
  {journal} {\bibinfo  {journal} {Phys. Rev. Lett.}\ }\textbf {\bibinfo
  {volume} {50}},\ \bibinfo {pages} {1395} (\bibinfo {year}
  {1983})}\BibitemShut {NoStop}%
\bibitem [{\citenamefont {Arovas}\ \emph {et~al.}(1984)\citenamefont {Arovas},
  \citenamefont {Schrieffer},\ and\ \citenamefont {Wilczek}}]{Arovas_84}%
  \BibitemOpen
  \bibfield  {author} {\bibinfo {author} {\bibfnamefont {D.}~\bibnamefont
  {Arovas}}, \bibinfo {author} {\bibfnamefont {J.~R.}\ \bibnamefont
  {Schrieffer}}, \ and\ \bibinfo {author} {\bibfnamefont {F.}~\bibnamefont
  {Wilczek}},\ }\href {\doibase 10.1103/PhysRevLett.53.722} {\bibfield
  {journal} {\bibinfo  {journal} {Phys. Rev. Lett.}\ }\textbf {\bibinfo
  {volume} {53}},\ \bibinfo {pages} {722} (\bibinfo {year} {1984})}\BibitemShut
  {NoStop}%
\bibitem [{\citenamefont {Thouless}\ \emph {et~al.}(1982)\citenamefont
  {Thouless}, \citenamefont {Kohmoto}, \citenamefont {Nightingale},\ and\
  \citenamefont {den Nijs}}]{PhysRevLett.49.405}%
  \BibitemOpen
  \bibfield  {author} {\bibinfo {author} {\bibfnamefont {D.~J.}\ \bibnamefont
  {Thouless}}, \bibinfo {author} {\bibfnamefont {M.}~\bibnamefont {Kohmoto}},
  \bibinfo {author} {\bibfnamefont {M.~P.}\ \bibnamefont {Nightingale}}, \ and\
  \bibinfo {author} {\bibfnamefont {M.}~\bibnamefont {den Nijs}},\ }\href
  {\doibase 10.1103/PhysRevLett.49.405} {\bibfield  {journal} {\bibinfo
  {journal} {Phys. Rev. Lett.}\ }\textbf {\bibinfo {volume} {49}},\ \bibinfo
  {pages} {405} (\bibinfo {year} {1982})}\BibitemShut {NoStop}%
\bibitem [{\citenamefont {Kane}\ and\ \citenamefont
  {Mele}(2005)}]{PhysRevLett.95.146802}%
  \BibitemOpen
  \bibfield  {author} {\bibinfo {author} {\bibfnamefont {C.~L.}\ \bibnamefont
  {Kane}}\ and\ \bibinfo {author} {\bibfnamefont {E.~J.}\ \bibnamefont
  {Mele}},\ }\href {\doibase 10.1103/PhysRevLett.95.146802} {\bibfield
  {journal} {\bibinfo  {journal} {Phys. Rev. Lett.}\ }\textbf {\bibinfo
  {volume} {95}},\ \bibinfo {pages} {146802} (\bibinfo {year}
  {2005})}\BibitemShut {NoStop}%
\bibitem [{\citenamefont {Fu}\ \emph {et~al.}(2007)\citenamefont {Fu},
  \citenamefont {Kane},\ and\ \citenamefont {Mele}}]{PhysRevLett.98.106803}%
  \BibitemOpen
  \bibfield  {author} {\bibinfo {author} {\bibfnamefont {L.}~\bibnamefont
  {Fu}}, \bibinfo {author} {\bibfnamefont {C.~L.}\ \bibnamefont {Kane}}, \ and\
  \bibinfo {author} {\bibfnamefont {E.~J.}\ \bibnamefont {Mele}},\ }\href
  {\doibase 10.1103/PhysRevLett.98.106803} {\bibfield  {journal} {\bibinfo
  {journal} {Phys. Rev. Lett.}\ }\textbf {\bibinfo {volume} {98}},\ \bibinfo
  {pages} {106803} (\bibinfo {year} {2007})}\BibitemShut {NoStop}%
\bibitem [{\citenamefont {Haldane}(1988)}]{PhysRevLett.61.2015}%
  \BibitemOpen
  \bibfield  {author} {\bibinfo {author} {\bibfnamefont {F.~D.~M.}\
  \bibnamefont {Haldane}},\ }\href {\doibase 10.1103/PhysRevLett.61.2015}
  {\bibfield  {journal} {\bibinfo  {journal} {Phys. Rev. Lett.}\ }\textbf
  {\bibinfo {volume} {61}},\ \bibinfo {pages} {2015} (\bibinfo {year}
  {1988})}\BibitemShut {NoStop}%
\bibitem [{\citenamefont {Lundholm}\ and\ \citenamefont
  {Rougerie}(2016)}]{Lundholm_2016}%
  \BibitemOpen
  \bibfield  {author} {\bibinfo {author} {\bibfnamefont {D.}~\bibnamefont
  {Lundholm}}\ and\ \bibinfo {author} {\bibfnamefont {N.}~\bibnamefont
  {Rougerie}},\ }\href {\doibase 10.1103/PhysRevLett.116.170401} {\bibfield
  {journal} {\bibinfo  {journal} {Phys. Rev. Lett.}\ }\textbf {\bibinfo
  {volume} {116}},\ \bibinfo {pages} {170401} (\bibinfo {year}
  {2016})}\BibitemShut {NoStop}%
\bibitem [{\citenamefont {Kitaev}(2003)}]{K}%
  \BibitemOpen
  \bibfield  {author} {\bibinfo {author} {\bibfnamefont {A.}~\bibnamefont
  {Kitaev}},\ }\href {\doibase 10.1016/s0003-4916(02)00018-0} {\bibfield
  {journal} {\bibinfo  {journal} {Annals of Physics}\ }\textbf {\bibinfo
  {volume} {303}},\ \bibinfo {pages} {2–30} (\bibinfo {year}
  {2003})}\BibitemShut {NoStop}%
\bibitem [{\citenamefont {Lloyd}(2002)}]{lloyd2002quantum}%
  \BibitemOpen
  \bibfield  {author} {\bibinfo {author} {\bibfnamefont {S.}~\bibnamefont
  {Lloyd}},\ }\href@noop {} {\bibfield  {journal} {\bibinfo  {journal} {Quantum
  Information Processing}\ }\textbf {\bibinfo {volume} {1}},\ \bibinfo {pages}
  {13} (\bibinfo {year} {2002})}\BibitemShut {NoStop}%
\bibitem [{\citenamefont {Freedman}\ \emph {et~al.}(2003)\citenamefont
  {Freedman}, \citenamefont {Kitaev}, \citenamefont {Larsen},\ and\
  \citenamefont {Wang}}]{freedman2003topological}%
  \BibitemOpen
  \bibfield  {author} {\bibinfo {author} {\bibfnamefont {M.}~\bibnamefont
  {Freedman}}, \bibinfo {author} {\bibfnamefont {A.}~\bibnamefont {Kitaev}},
  \bibinfo {author} {\bibfnamefont {M.}~\bibnamefont {Larsen}}, \ and\ \bibinfo
  {author} {\bibfnamefont {Z.}~\bibnamefont {Wang}},\ }\href@noop {} {\bibfield
   {journal} {\bibinfo  {journal} {Bulletin of the American Mathematical
  Society}\ }\textbf {\bibinfo {volume} {40}},\ \bibinfo {pages} {31} (\bibinfo
  {year} {2003})}\BibitemShut {NoStop}%
\bibitem [{\citenamefont {Nayak}\ \emph {et~al.}(2008)\citenamefont {Nayak},
  \citenamefont {Simon}, \citenamefont {Stern}, \citenamefont {Freedman},\ and\
  \citenamefont {Das~Sarma}}]{Nayak_08}%
  \BibitemOpen
  \bibfield  {author} {\bibinfo {author} {\bibfnamefont {C.}~\bibnamefont
  {Nayak}}, \bibinfo {author} {\bibfnamefont {S.~H.}\ \bibnamefont {Simon}},
  \bibinfo {author} {\bibfnamefont {A.}~\bibnamefont {Stern}}, \bibinfo
  {author} {\bibfnamefont {M.}~\bibnamefont {Freedman}}, \ and\ \bibinfo
  {author} {\bibfnamefont {S.}~\bibnamefont {Das~Sarma}},\ }\href {\doibase
  10.1103/RevModPhys.80.1083} {\bibfield  {journal} {\bibinfo  {journal} {Rev.
  Mod. Phys.}\ }\textbf {\bibinfo {volume} {80}},\ \bibinfo {pages} {1083}
  (\bibinfo {year} {2008})}\BibitemShut {NoStop}%
\bibitem [{\citenamefont {Leinaas}\ and\ \citenamefont
  {Myrheim}(1977)}]{leinaas1977theory}%
  \BibitemOpen
  \bibfield  {author} {\bibinfo {author} {\bibfnamefont {J.~M.}\ \bibnamefont
  {Leinaas}}\ and\ \bibinfo {author} {\bibfnamefont {J.}~\bibnamefont
  {Myrheim}},\ }\href@noop {} {\bibfield  {journal} {\bibinfo  {journal} {Il
  Nuovo Cimento B (1971-1996)}\ }\textbf {\bibinfo {volume} {37}},\ \bibinfo
  {pages} {1} (\bibinfo {year} {1977})}\BibitemShut {NoStop}%
\bibitem [{\citenamefont {Wilczek}(1982{\natexlab{a}})}]{Wilczek_82}%
  \BibitemOpen
  \bibfield  {author} {\bibinfo {author} {\bibfnamefont {F.}~\bibnamefont
  {Wilczek}},\ }\href {\doibase 10.1103/PhysRevLett.48.1144} {\bibfield
  {journal} {\bibinfo  {journal} {Phys. Rev. Lett.}\ }\textbf {\bibinfo
  {volume} {48}},\ \bibinfo {pages} {1144} (\bibinfo {year}
  {1982}{\natexlab{a}})}\BibitemShut {NoStop}%
\bibitem [{\citenamefont {Wilczek}(1982{\natexlab{b}})}]{Wilczek_82b}%
  \BibitemOpen
  \bibfield  {author} {\bibinfo {author} {\bibfnamefont {F.}~\bibnamefont
  {Wilczek}},\ }\href {\doibase 10.1103/PhysRevLett.49.957} {\bibfield
  {journal} {\bibinfo  {journal} {Phys. Rev. Lett.}\ }\textbf {\bibinfo
  {volume} {49}},\ \bibinfo {pages} {957} (\bibinfo {year}
  {1982}{\natexlab{b}})}\BibitemShut {NoStop}%
\bibitem [{\citenamefont {Goldin}\ \emph {et~al.}(1981)\citenamefont {Goldin},
  \citenamefont {Menikoff},\ and\ \citenamefont
  {Sharp}}]{goldin1981representations}%
  \BibitemOpen
  \bibfield  {author} {\bibinfo {author} {\bibfnamefont {G.~A.}\ \bibnamefont
  {Goldin}}, \bibinfo {author} {\bibfnamefont {R.}~\bibnamefont {Menikoff}}, \
  and\ \bibinfo {author} {\bibfnamefont {D.~H.}\ \bibnamefont {Sharp}},\
  }\href@noop {} {\bibfield  {journal} {\bibinfo  {journal} {Journal of
  Mathematical Physics}\ }\textbf {\bibinfo {volume} {22}},\ \bibinfo {pages}
  {1664} (\bibinfo {year} {1981})}\BibitemShut {NoStop}%
\bibitem [{\citenamefont {Wu}(1984)}]{Wu_84}%
  \BibitemOpen
  \bibfield  {author} {\bibinfo {author} {\bibfnamefont {Y.-S.}\ \bibnamefont
  {Wu}},\ }\href {\doibase 10.1103/PhysRevLett.52.2103} {\bibfield  {journal}
  {\bibinfo  {journal} {Phys. Rev. Lett.}\ }\textbf {\bibinfo {volume} {52}},\
  \bibinfo {pages} {2103} (\bibinfo {year} {1984})}\BibitemShut {NoStop}%
\bibitem [{\citenamefont {Cooper}\ and\ \citenamefont
  {Simon}(2015)}]{CooSim-15}%
  \BibitemOpen
  \bibfield  {author} {\bibinfo {author} {\bibfnamefont {N.~R.}\ \bibnamefont
  {Cooper}}\ and\ \bibinfo {author} {\bibfnamefont {S.~H.}\ \bibnamefont
  {Simon}},\ }\href {\doibase 10.1103/PhysRevLett.114.106802} {\bibfield
  {journal} {\bibinfo  {journal} {Phys. Rev. Lett.}\ }\textbf {\bibinfo
  {volume} {114}},\ \bibinfo {pages} {106802} (\bibinfo {year}
  {2015})}\BibitemShut {NoStop}%
\bibitem [{\citenamefont {Zhang}\ \emph {et~al.}(2014)\citenamefont {Zhang},
  \citenamefont {Sreejith}, \citenamefont {Gemelke},\ and\ \citenamefont
  {Jain}}]{ZhaSreGemJai-14}%
  \BibitemOpen
  \bibfield  {author} {\bibinfo {author} {\bibfnamefont {Y.}~\bibnamefont
  {Zhang}}, \bibinfo {author} {\bibfnamefont {G.~J.}\ \bibnamefont {Sreejith}},
  \bibinfo {author} {\bibfnamefont {N.~D.}\ \bibnamefont {Gemelke}}, \ and\
  \bibinfo {author} {\bibfnamefont {J.~K.}\ \bibnamefont {Jain}},\ }\href
  {\doibase 10.1103/PhysRevLett.113.160404} {\bibfield  {journal} {\bibinfo
  {journal} {Phys. Rev. Lett.}\ }\textbf {\bibinfo {volume} {113}},\ \bibinfo
  {pages} {160404} (\bibinfo {year} {2014})}\BibitemShut {NoStop}%
\bibitem [{\citenamefont {Zhang}\ \emph {et~al.}(2015)\citenamefont {Zhang},
  \citenamefont {Sreejith},\ and\ \citenamefont {Jain}}]{ZhaSreJai-15}%
  \BibitemOpen
  \bibfield  {author} {\bibinfo {author} {\bibfnamefont {Y.}~\bibnamefont
  {Zhang}}, \bibinfo {author} {\bibfnamefont {G.~J.}\ \bibnamefont {Sreejith}},
  \ and\ \bibinfo {author} {\bibfnamefont {J.~K.}\ \bibnamefont {Jain}},\
  }\href {\doibase 10.1103/PhysRevB.92.075116} {\bibfield  {journal} {\bibinfo
  {journal} {Phys. Rev. B.}\ }\textbf {\bibinfo {volume} {92}},\ \bibinfo
  {pages} {075116} (\bibinfo {year} {2015})}\BibitemShut {NoStop}%
\bibitem [{\citenamefont {Morampudi}\ \emph {et~al.}(2017)\citenamefont
  {Morampudi}, \citenamefont {Turner}, \citenamefont {Pollmann},\ and\
  \citenamefont {Wilczek}}]{MorTurPolWil-17}%
  \BibitemOpen
  \bibfield  {author} {\bibinfo {author} {\bibfnamefont {S.~C.}\ \bibnamefont
  {Morampudi}}, \bibinfo {author} {\bibfnamefont {A.~M.}\ \bibnamefont
  {Turner}}, \bibinfo {author} {\bibfnamefont {F.}~\bibnamefont {Pollmann}}, \
  and\ \bibinfo {author} {\bibfnamefont {F.}~\bibnamefont {Wilczek}},\ }\href
  {\doibase 10.1103/PhysRevLett.118.227201} {\bibfield  {journal} {\bibinfo
  {journal} {Phys. Rev. Lett.}\ }\textbf {\bibinfo {volume} {118}},\ \bibinfo
  {pages} {227201} (\bibinfo {year} {2017})}\BibitemShut {NoStop}%
\bibitem [{\citenamefont {Umucal\ifmmode\imath\else\i\fi{}lar}\ \emph
  {et~al.}(2018)\citenamefont {Umucal\ifmmode\imath\else\i\fi{}lar},
  \citenamefont {Macaluso}, \citenamefont {Comparin},\ and\ \citenamefont
  {Carusotto}}]{UmuMacComCar-18}%
  \BibitemOpen
  \bibfield  {author} {\bibinfo {author} {\bibfnamefont {R.~O.}\ \bibnamefont
  {Umucal\ifmmode\imath\else\i\fi{}lar}}, \bibinfo {author} {\bibfnamefont
  {E.}~\bibnamefont {Macaluso}}, \bibinfo {author} {\bibfnamefont
  {T.}~\bibnamefont {Comparin}}, \ and\ \bibinfo {author} {\bibfnamefont
  {I.}~\bibnamefont {Carusotto}},\ }\href {\doibase
  10.1103/PhysRevLett.120.230403} {\bibfield  {journal} {\bibinfo  {journal}
  {Phys. Rev. Lett.}\ }\textbf {\bibinfo {volume} {120}},\ \bibinfo {pages}
  {230403} (\bibinfo {year} {2018})}\BibitemShut {NoStop}%
\bibitem [{\citenamefont {Correggi}\ \emph {et~al.}(2019)\citenamefont
  {Correggi}, \citenamefont {Duboscq}, \citenamefont {Lundholm},\ and\
  \citenamefont {Rougerie}}]{CorDubLunRou-19}%
  \BibitemOpen
  \bibfield  {author} {\bibinfo {author} {\bibfnamefont {M.}~\bibnamefont
  {Correggi}}, \bibinfo {author} {\bibfnamefont {R.}~\bibnamefont {Duboscq}},
  \bibinfo {author} {\bibfnamefont {D.}~\bibnamefont {Lundholm}}, \ and\
  \bibinfo {author} {\bibfnamefont {N.}~\bibnamefont {Rougerie}},\ }\href
  {\doibase 10.1209/0295-5075/126/20005} {\bibfield  {journal} {\bibinfo
  {journal} {EPL (Europhysics Letters)}\ }\textbf {\bibinfo {volume} {126}},\
  \bibinfo {pages} {20005} (\bibinfo {year} {2019})}\BibitemShut {NoStop}%
\bibitem [{\citenamefont {Yakaboylu}\ and\ \citenamefont
  {Lemeshko}(2018)}]{Yakaboylu_2018}%
  \BibitemOpen
  \bibfield  {author} {\bibinfo {author} {\bibfnamefont {E.}~\bibnamefont
  {Yakaboylu}}\ and\ \bibinfo {author} {\bibfnamefont {M.}~\bibnamefont
  {Lemeshko}},\ }\href {\doibase 10.1103/PhysRevB.98.045402} {\bibfield
  {journal} {\bibinfo  {journal} {Phys. Rev. B}\ }\textbf {\bibinfo {volume}
  {98}},\ \bibinfo {pages} {045402} (\bibinfo {year} {2018})}\BibitemShut
  {NoStop}%
\bibitem [{\citenamefont {Yakaboylu}\ \emph {et~al.}(2020)\citenamefont
  {Yakaboylu}, \citenamefont {Ghazaryan}, \citenamefont {Lundholm},
  \citenamefont {Rougerie}, \citenamefont {Lemeshko},\ and\ \citenamefont
  {Seiringer}}]{Yakaboylu_anyon_2020}%
  \BibitemOpen
  \bibfield  {author} {\bibinfo {author} {\bibfnamefont {E.}~\bibnamefont
  {Yakaboylu}}, \bibinfo {author} {\bibfnamefont {A.}~\bibnamefont
  {Ghazaryan}}, \bibinfo {author} {\bibfnamefont {D.}~\bibnamefont {Lundholm}},
  \bibinfo {author} {\bibfnamefont {N.}~\bibnamefont {Rougerie}}, \bibinfo
  {author} {\bibfnamefont {M.}~\bibnamefont {Lemeshko}}, \ and\ \bibinfo
  {author} {\bibfnamefont {R.}~\bibnamefont {Seiringer}},\ }\href {\doibase
  10.1103/PhysRevB.102.144109} {\bibfield  {journal} {\bibinfo  {journal}
  {Phys. Rev. B}\ }\textbf {\bibinfo {volume} {102}},\ \bibinfo {pages}
  {144109} (\bibinfo {year} {2020})}\BibitemShut {NoStop}%
\bibitem [{\citenamefont {Thouless}\ and\ \citenamefont
  {Wu}(1985)}]{Thouless_85}%
  \BibitemOpen
  \bibfield  {author} {\bibinfo {author} {\bibfnamefont {D.~J.}\ \bibnamefont
  {Thouless}}\ and\ \bibinfo {author} {\bibfnamefont {Y.-S.}\ \bibnamefont
  {Wu}},\ }\href {\doibase 10.1103/PhysRevB.31.1191} {\bibfield  {journal}
  {\bibinfo  {journal} {Phys. Rev. B}\ }\textbf {\bibinfo {volume} {31}},\
  \bibinfo {pages} {1191} (\bibinfo {year} {1985})}\BibitemShut {NoStop}%
\bibitem [{\citenamefont {Einarsson}(1990)}]{Einarsson_90}%
  \BibitemOpen
  \bibfield  {author} {\bibinfo {author} {\bibfnamefont {T.}~\bibnamefont
  {Einarsson}},\ }\href {\doibase 10.1103/PhysRevLett.64.1995} {\bibfield
  {journal} {\bibinfo  {journal} {Phys. Rev. Lett.}\ }\textbf {\bibinfo
  {volume} {64}},\ \bibinfo {pages} {1995} (\bibinfo {year}
  {1990})}\BibitemShut {NoStop}%
\bibitem [{\citenamefont {Comtet}\ \emph {et~al.}(1992)\citenamefont {Comtet},
  \citenamefont {McCabe},\ and\ \citenamefont {Ouvry}}]{CMO}%
  \BibitemOpen
  \bibfield  {author} {\bibinfo {author} {\bibfnamefont {A.}~\bibnamefont
  {Comtet}}, \bibinfo {author} {\bibfnamefont {J.}~\bibnamefont {McCabe}}, \
  and\ \bibinfo {author} {\bibfnamefont {S.}~\bibnamefont {Ouvry}},\ }\href
  {\doibase 10.1103/PhysRevD.45.709} {\bibfield  {journal} {\bibinfo  {journal}
  {Phys. Rev. D}\ }\textbf {\bibinfo {volume} {45}},\ \bibinfo {pages} {709}
  (\bibinfo {year} {1992})}\BibitemShut {NoStop}%
\bibitem [{\citenamefont {Einarsson}(1991)}]{einarsson1991fractional}%
  \BibitemOpen
  \bibfield  {author} {\bibinfo {author} {\bibfnamefont {T.}~\bibnamefont
  {Einarsson}},\ }\href@noop {} {\bibfield  {journal} {\bibinfo  {journal}
  {Modern Physics Letters B}\ }\textbf {\bibinfo {volume} {5}},\ \bibinfo
  {pages} {675} (\bibinfo {year} {1991})}\BibitemShut {NoStop}%
\bibitem [{\citenamefont {Ouvry}\ and\ \citenamefont
  {Polychronakos}(2019)}]{ouvry2019anyons}%
  \BibitemOpen
  \bibfield  {author} {\bibinfo {author} {\bibfnamefont {S.}~\bibnamefont
  {Ouvry}}\ and\ \bibinfo {author} {\bibfnamefont {A.~P.}\ \bibnamefont
  {Polychronakos}},\ }\href@noop {} {\bibfield  {journal} {\bibinfo  {journal}
  {Nuclear Physics B}\ }\textbf {\bibinfo {volume} {949}},\ \bibinfo {pages}
  {114797} (\bibinfo {year} {2019})}\BibitemShut {NoStop}%
\bibitem [{\citenamefont {Polychronakos}\ and\ \citenamefont
  {Ouvry}(2020)}]{polychronakos2020two}%
  \BibitemOpen
  \bibfield  {author} {\bibinfo {author} {\bibfnamefont {A.~P.}\ \bibnamefont
  {Polychronakos}}\ and\ \bibinfo {author} {\bibfnamefont {S.}~\bibnamefont
  {Ouvry}},\ }\href@noop {} {\bibfield  {journal} {\bibinfo  {journal} {Nuclear
  Physics B}\ }\textbf {\bibinfo {volume} {951}},\ \bibinfo {pages} {114906}
  (\bibinfo {year} {2020})}\BibitemShut {NoStop}%
\bibitem [{\citenamefont {Haldane}(1983)}]{Haldane_83}%
  \BibitemOpen
  \bibfield  {author} {\bibinfo {author} {\bibfnamefont {F.~D.~M.}\
  \bibnamefont {Haldane}},\ }\href {\doibase 10.1103/PhysRevLett.51.605}
  {\bibfield  {journal} {\bibinfo  {journal} {Phys. Rev. Lett.}\ }\textbf
  {\bibinfo {volume} {51}},\ \bibinfo {pages} {605} (\bibinfo {year}
  {1983})}\BibitemShut {NoStop}%
\bibitem [{\citenamefont {Harrison}\ \emph {et~al.}(2014)\citenamefont
  {Harrison}, \citenamefont {Keating}, \citenamefont {Robbins},\ and\
  \citenamefont {Sawicki}}]{Harrison-etal-14}%
  \BibitemOpen
  \bibfield  {author} {\bibinfo {author} {\bibfnamefont {J.~M.}\ \bibnamefont
  {Harrison}}, \bibinfo {author} {\bibfnamefont {J.~P.}\ \bibnamefont
  {Keating}}, \bibinfo {author} {\bibfnamefont {J.~M.}\ \bibnamefont
  {Robbins}}, \ and\ \bibinfo {author} {\bibfnamefont {A.}~\bibnamefont
  {Sawicki}},\ }\href {\doibase 10.1007/s00220-014-2091-0} {\bibfield
  {journal} {\bibinfo  {journal} {Comm. Math. Phys.}\ }\textbf {\bibinfo
  {volume} {330}},\ \bibinfo {pages} {1293} (\bibinfo {year}
  {2014})}\BibitemShut {NoStop}%
\bibitem [{\citenamefont {Maciazek}\ and\ \citenamefont
  {Sawicki}(2019)}]{MacSaw-19}%
  \BibitemOpen
  \bibfield  {author} {\bibinfo {author} {\bibfnamefont {T.}~\bibnamefont
  {Maciazek}}\ and\ \bibinfo {author} {\bibfnamefont {A.}~\bibnamefont
  {Sawicki}},\ }\href {\doibase 10.1007/s00220-019-03583-5} {\bibfield
  {journal} {\bibinfo  {journal} {Commun. Math. Phys.}\ }\textbf {\bibinfo
  {volume} {371}},\ \bibinfo {pages} {921} (\bibinfo {year}
  {2019})}\BibitemShut {NoStop}%
\bibitem [{\citenamefont {Brooks}\ \emph {et~al.}(2021)\citenamefont {Brooks},
  \citenamefont {Lemeshko}, \citenamefont {Lundholm},\ and\ \citenamefont
  {Yakaboylu}}]{Brooks_2021}%
  \BibitemOpen
  \bibfield  {author} {\bibinfo {author} {\bibfnamefont {M.}~\bibnamefont
  {Brooks}}, \bibinfo {author} {\bibfnamefont {M.}~\bibnamefont {Lemeshko}},
  \bibinfo {author} {\bibfnamefont {D.}~\bibnamefont {Lundholm}}, \ and\
  \bibinfo {author} {\bibfnamefont {E.}~\bibnamefont {Yakaboylu}},\ }\href
  {\doibase 10.1103/PhysRevLett.126.015301} {\bibfield  {journal} {\bibinfo
  {journal} {Phys. Rev. Lett.}\ }\textbf {\bibinfo {volume} {126}},\ \bibinfo
  {pages} {015301} (\bibinfo {year} {2021})}\BibitemShut {NoStop}%
\bibitem [{\citenamefont {Schmidt}\ and\ \citenamefont
  {Lemeshko}(2015)}]{Lemeshko_2015}%
  \BibitemOpen
  \bibfield  {author} {\bibinfo {author} {\bibfnamefont {R.}~\bibnamefont
  {Schmidt}}\ and\ \bibinfo {author} {\bibfnamefont {M.}~\bibnamefont
  {Lemeshko}},\ }\href {\doibase 10.1103/PhysRevLett.114.203001} {\bibfield
  {journal} {\bibinfo  {journal} {Phys. Rev. Lett.}\ }\textbf {\bibinfo
  {volume} {114}},\ \bibinfo {pages} {203001} (\bibinfo {year}
  {2015})}\BibitemShut {NoStop}%
\bibitem [{\citenamefont {Schmidt}\ and\ \citenamefont
  {Lemeshko}(2016)}]{PhysRevX.6.011012}%
  \BibitemOpen
  \bibfield  {author} {\bibinfo {author} {\bibfnamefont {R.}~\bibnamefont
  {Schmidt}}\ and\ \bibinfo {author} {\bibfnamefont {M.}~\bibnamefont
  {Lemeshko}},\ }\href {\doibase 10.1103/PhysRevX.6.011012} {\bibfield
  {journal} {\bibinfo  {journal} {Phys. Rev. X}\ }\textbf {\bibinfo {volume}
  {6}},\ \bibinfo {pages} {011012} (\bibinfo {year} {2016})}\BibitemShut
  {NoStop}%
\bibitem [{\citenamefont {Lemeshko}\ and\ \citenamefont
  {Schmidt}(2016)}]{Lemeshko_2016_book}%
  \BibitemOpen
  \bibfield  {author} {\bibinfo {author} {\bibfnamefont {M.}~\bibnamefont
  {Lemeshko}}\ and\ \bibinfo {author} {\bibfnamefont {R.}~\bibnamefont
  {Schmidt}},\ }\href@noop {} {\emph {\bibinfo {title} {Molecular impurities
  interacting with a many-particle environment: from ultracold gases to helium
  nanodroplets, book chapter in "Low Energy and Low Temperature Molecular
  Scattering" edited by A. Osterwalder and O. Dulieu}}}\ (\bibinfo  {publisher}
  {Royal Society of Chemistry},\ \bibinfo {year} {2016})\BibitemShut {NoStop}%
\bibitem [{\citenamefont {Lemeshko}(2017)}]{lemeshko2016quasiparticle}%
  \BibitemOpen
  \bibfield  {author} {\bibinfo {author} {\bibfnamefont {M.}~\bibnamefont
  {Lemeshko}},\ }\href {\doibase 10.1103/PhysRevLett.118.095301} {\bibfield
  {journal} {\bibinfo  {journal} {Phys. Rev. Lett.}\ }\textbf {\bibinfo
  {volume} {118}},\ \bibinfo {pages} {095301} (\bibinfo {year}
  {2017})}\BibitemShut {NoStop}%
\bibitem [{\citenamefont {Shchadilova}(2017)}]{YuliaPhysics17}%
  \BibitemOpen
  \bibfield  {author} {\bibinfo {author} {\bibfnamefont {Y.}~\bibnamefont
  {Shchadilova}},\ }\href@noop {} {\bibfield  {journal} {\bibinfo  {journal}
  {Physics}\ }\textbf {\bibinfo {volume} {10}},\ \bibinfo {pages} {20}
  (\bibinfo {year} {2017})}\BibitemShut {NoStop}%
\bibitem [{\citenamefont {Shepperson}\ \emph
  {et~al.}(2017{\natexlab{a}})\citenamefont {Shepperson}, \citenamefont
  {S{\"o}ndergaard}, \citenamefont {Christiansen}, \citenamefont {Kaczmarczyk},
  \citenamefont {Zillich}, \citenamefont {Lemeshko},\ and\ \citenamefont
  {Stapelfeldt}}]{Shepperson16}%
  \BibitemOpen
  \bibfield  {author} {\bibinfo {author} {\bibfnamefont {B.}~\bibnamefont
  {Shepperson}}, \bibinfo {author} {\bibfnamefont {A.~A.}\ \bibnamefont
  {S{\"o}ndergaard}}, \bibinfo {author} {\bibfnamefont {L.}~\bibnamefont
  {Christiansen}}, \bibinfo {author} {\bibfnamefont {J.}~\bibnamefont
  {Kaczmarczyk}}, \bibinfo {author} {\bibfnamefont {R.~E.}\ \bibnamefont
  {Zillich}}, \bibinfo {author} {\bibfnamefont {M.}~\bibnamefont {Lemeshko}}, \
  and\ \bibinfo {author} {\bibfnamefont {H.}~\bibnamefont {Stapelfeldt}},\
  }\href@noop {} {\bibfield  {journal} {\bibinfo  {journal} {submitted}\ }
  (\bibinfo {year} {2017}{\natexlab{a}})}\BibitemShut {NoStop}%
\bibitem [{\citenamefont {Shepperson}\ \emph
  {et~al.}(2017{\natexlab{b}})\citenamefont {Shepperson}, \citenamefont
  {Chatterley}, \citenamefont {S{\o}ndergaard}, \citenamefont {Christiansen},
  \citenamefont {Lemeshko},\ and\ \citenamefont {Stapelfeldt}}]{Shepperson17}%
  \BibitemOpen
  \bibfield  {author} {\bibinfo {author} {\bibfnamefont {B.}~\bibnamefont
  {Shepperson}}, \bibinfo {author} {\bibfnamefont {A.~S.}\ \bibnamefont
  {Chatterley}}, \bibinfo {author} {\bibfnamefont {A.~A.}\ \bibnamefont
  {S{\o}ndergaard}}, \bibinfo {author} {\bibfnamefont {L.}~\bibnamefont
  {Christiansen}}, \bibinfo {author} {\bibfnamefont {M.}~\bibnamefont
  {Lemeshko}}, \ and\ \bibinfo {author} {\bibfnamefont {H.}~\bibnamefont
  {Stapelfeldt}},\ }\href@noop {} {\bibfield  {journal} {\bibinfo  {journal}
  {J. Chem. Phys. (in press); arXiv:1704.03684}\ } (\bibinfo {year}
  {2017}{\natexlab{b}})}\BibitemShut {NoStop}%
\bibitem [{\citenamefont {Bourdeau}\ and\ \citenamefont
  {Sorkin}(1992)}]{BorSor-92}%
  \BibitemOpen
  \bibfield  {author} {\bibinfo {author} {\bibfnamefont {M.}~\bibnamefont
  {Bourdeau}}\ and\ \bibinfo {author} {\bibfnamefont {R.~D.}\ \bibnamefont
  {Sorkin}},\ }\href {\doibase 10.1103/PhysRevD.45.687} {\bibfield  {journal}
  {\bibinfo  {journal} {Phys. Rev. D}\ }\textbf {\bibinfo {volume} {45}},\
  \bibinfo {pages} {687} (\bibinfo {year} {1992})}\BibitemShut {NoStop}%
\bibitem [{\citenamefont {Lundholm}\ and\ \citenamefont
  {Solovej}(2014)}]{LunSol-14}%
  \BibitemOpen
  \bibfield  {author} {\bibinfo {author} {\bibfnamefont {D.}~\bibnamefont
  {Lundholm}}\ and\ \bibinfo {author} {\bibfnamefont {J.~P.}\ \bibnamefont
  {Solovej}},\ }\href {\doibase 10.1007/s00023-013-0273-5} {\bibfield
  {journal} {\bibinfo  {journal} {Ann. Henri Poincar\'e}\ }\textbf {\bibinfo
  {volume} {15}},\ \bibinfo {pages} {1061} (\bibinfo {year}
  {2014})}\BibitemShut {NoStop}%
\bibitem [{\citenamefont {Correggi}\ and\ \citenamefont
  {Oddis}(2018)}]{CorOdd-18}%
  \BibitemOpen
  \bibfield  {author} {\bibinfo {author} {\bibfnamefont {M.}~\bibnamefont
  {Correggi}}\ and\ \bibinfo {author} {\bibfnamefont {L.}~\bibnamefont
  {Oddis}},\ }\href
  {http://www1.mat.uniroma1.it/ricerca/rendiconti/39_2_(2018)_277-292.html}
  {\bibfield  {journal} {\bibinfo  {journal} {Rend. Mat. Appl.}\ }\textbf
  {\bibinfo {volume} {39}},\ \bibinfo {pages} {277} (\bibinfo {year}
  {2018})}\BibitemShut {NoStop}%
\end{thebibliography}%

\end{document}